# Substantial Doubt Remains about the Efficacy of Anti-Amyloid Antibodies

Leonardino A. Digma, MD[1], Joseph R. Winer, PhD[1], Michael D. Greicius, MD[1]

[1]Department of Neurology and Neurological Sciences, Stanford University, Stanford, California

Running Title: Doubt remains about anti-amyloid antibodies
Word Count: 2057
Number of Figures: 2

**Corresponding Author:**
Dr. Michael D. Greicius
Stanford Neuroscience Health Center
290 Jane Stanford Way
Stanford, CA 94305-5090
greicius@stanford.edu
(650) 498-4624

## Abstract

With the FDA approval of aducanumab and lecanemab, and with the recent statistically significant phase 3 clinical trial for donanemab, there is growing enthusiasm for anti-amyloid antibodies in the treatment of Alzheimer's disease. Here, we discuss three substantial limitations regarding recent anti-amyloid clinical trials: (1) there is little evidence that amyloid reduction correlates with clinical outcome, (2) the reported efficacy of anti-amyloid therapies may be explained by functional unblinding, and (3) donanemab had no effect on tau burden in its phase 3 trial. Taken together, these observations call into question the efficacy of anti-amyloid therapies.

## Introduction

Over the past fifteen years, several antibodies designed to reduce brain amyloid have made their way into phase 3 clinical trials of Alzheimer's disease (AD). Since 2021, two such therapies, aducanumab and lecanemab, have been approved by the FDA[1,2] for use in MCI due to AD and mild AD. New anti-amyloid therapies continue to progress through the pipeline and, most recently, a phase 3 clinical trial of donanemab generated a statistically significant outcome[3]. While monoclonal antibodies have not yet become routine in clinical practice, the recent series of statistically positive trials has generated enthusiasm among some, though by no means all, AD experts [4–7]. As the field enters a new era of amyloid-modifying therapies we highlight three substantial limitations of these trials that cast doubt on the clinical efficacy of anti-amyloid antibodies.

## Amyloid Plaque Burden Is Not a Useful Biomarker for Predicting Clinical Efficacy

Under the FDA Accelerated Approval pathway, potential therapies for conditions with an unmet need can be approved based on a surrogate endpoint (e.g., laboratory or imaging marker) that is likely to predict clinical benefit [8]. In the context of AD, reduction in brain amyloid has been suggested as a surrogate endpoint and it was under this specification that lecanemab and aducanumab received approval [1,2]. Lecanemab has since also received traditional approval [9]. There is little question that some species of the β-amyloid protein plays an early, crucial role in AD pathogenesis. There is little to no evidence, however, that the amount or location of amyloid plaque has any bearing on cognitive function and, critically, little to no evidence that the amount of plaque removed correlates with clinical outcomes. Long before amyloid PET, post-mortem studies demonstrated that tau pathology, measured by neurofibrillary tangle (NFT) burden, correlated regionally with neuronal loss and globally with cognitive status whereas amyloid plaque burden generally had little to no correlation locally or globally [10,11]. Multivariate approaches have shown that even when an association is found between amyloid plaque burden and global cognition the effect is attenuated or lost when NFT burden is in the model [12]. This lesson—NFTs, but not amyloid plaques, track regionally with impaired function and globally with cognitive decline—has been relearned in the era of amyloid and tau PET imaging [13,14]. As such, the FDA's decision to accept amyloid plaque reduction as a surrogate marker "reasonably likely to predict a clinical benefit" was met with skepticism [6].

The results of anti-amyloid antibody trials have justified this skepticism. We are only aware of a single instance, discussed below, in which a clinical trial reported a correlation between amyloid plaque removal and the primary clinical outcome. What is often presented, instead, is some version of Figure 1A, showing a correlation *across* studies rather than across participants *within* a study [15]. We asked Biogen, Eisai, and Lilly for access to the relevant data from the phase 3 studies of aducanumab, lecanemab, and donanemab but our requests were declined. The only subject-level within-trial data we could access was found in the "Statistical Review and Evaluation" from the dissenting FDA statisticians in the aducanumab briefing document. We extracted these data (available for high-dose arms) for the EMERGE and ENGAGE aducanumab trials from the briefing document and show the correlation plots in Figure 1B. In neither trial was there a significant correlation between these two measures. We expect the same is true for lecanemab and donanemab since positive correlations between amyloid reduction and clinical outcomes, would likely have been featured in the phase 3 publications. We cannot confirm this, however, as these data were not made available to us. The *only* instance in which a potential *within* trial association has been demonstrated is for EMERGE, where a retrospective analysis that arbitrarily pooled high- and low-dose groups showed a weak correlation between amyloid reduction and clinical outcome (Spearman r =0.19) [16] (See their supplementary Figure 4c). The limitations of this analysis have been discussed previously [6]. Medicine has numerous examples of useful biomarkers that change predictably in the setting of disease progression or treatment response. Viral load in HIV is a prime example of a biomarker likely to predict clinical benefit: The risk of opportunistic infections increases with increasing viral load and drugs that reduce viral load are associated with reduced morbidity and mortality[17]. Postmortem studies, amyloid PET studies, and now clinical trial results all provide converging evidence that amyloid plaque burden is not suitable as a surrogate biomarker in clinical trials of AD.

**Functional Unblinding Due to ARIA May Contribute Substantially to Clinical Outcomes**
In the phase 3 trials of aducanumab, lecanemab, and donanemab a sizable percentage (ranging from 21 to 44%) of active treatment participants experience brain swelling or brain hemorrhaging. This occasionally fatal side effect [18,19] has been given the soothing acronym ARIA for amyloid-related imaging abnormality (with E or H added to denote edema or hemorrhage). In addition to safety risks, ARIA poses a significant challenge to interpreting the outcome data. Most ARIA cases are asymptomatic and picked up on routine safety imaging but this adverse event is nonetheless likely to result in functional unblinding. Typically, when a patient develops ARIA the dosing is halted and/or the patient is asked to undergo more frequent MRI scans until the ARIA resolves. These protocol changes alert the patient, their informant, their physician, and potentially other study personnel to the fact that something abnormal was seen on a safety MRI. ARIA has been recognized as a common side effect in these trials since 2009 [20], so halting treatment and/or increasing MRI surveillance is a strong indicator that a participant is on active treatment. Given that the Clinical Dementia Rating Scale Sum of Boxes and similar measures have a sizable subjective component (including an interview with the patient's informant), there is a real risk for functional unblinding to bias outcomes. Even objective test components may be affected if we assume that patients themselves, particularly those with only mild impairment, are prone to a placebo effect when they suspect they are on active treatment.   This raises the critical question of how much of the clinical effect reported in these trials can be attributed to inadequate consideration of functional unblinding.

The effect sizes reported are already small and of dubious clinical relevance[21], but the true, biologic effect sizes are likely smaller still and, possibly, non-existent once functional unblinding is adequately addressed. Attempts to account for functional unblinding due to ARIA have been inadequate. In the recently published donanemab trial the investigators attempted to account for ARIA-E by performing a sensitivity analysis. Study participants that experienced ARIA-E essentially had their true post-ARIA data points replaced with extrapolated data points derived from the average clinical course of participants in the same study arm who did not get ARIA. It is well-established that patients with one or two copies of the e4 allele of the *APOE* gene are more likely to develop ARIA-E [22]. It is also well-established that *APOE* e4 carriers have a steeper longitudinal decline in cognition than e4 non-carriers [23,24]. Thus, in the sensitivity analysis, the apparent beneficial effect of donanemab may be due to the fact that the post-ARIA trajectory of a cohort enriched for e4 carriers (known to have worse cognitive trajectories) is being replaced with a longitudinal trajectory derived from a cohort of subjects enriched for non-e4 carriers (known to have better cognitive trajectories). Only the sparsest description of the sensitivity analysis employed in the lecanemab study was provided, mentioning that the primary mixed model with repeated measures was done with "repeated censoring assessments after occurrence of ARIA-E (modified intention-to-treat)." It is challenging to determine just how this was done. It seems the post-ARIA time points were censored which would result in the same flaw that hobbles the donanemab sensitivity analysis, namely fewer longitudinal datapoints for e4 carriers in the treatment group compared to the placebo group.

How could companies reassure prescribers of these costly and potentially dangerous drugs that functional unblinding did not impact outcomes? One approach would be to perform the same sensitivity analyses but matching ARIA patients in the active arm with age- and e4-matched patients in the placebo arm and performing the truncation and extrapolation steps similarly in each arm. It would also be exceedingly helpful to see longitudinal "spaghetti" plots of participant trajectories with the occurrence of ARIA timestamped. This would help determine if there is a post-ARIA "bump" in the outcome measure. It would also be helpful to see, within the treatment arm, how patients who developed ARIA fared compared to age- and e4 dose-matched subjects who did not develop ARIA. As we were not granted access to the necessary subject-level, *within*-trial data and could not find it in any FDA documents, we sought to determine *across* trials if the percentage of patients with ARIA correlated with clinical outcome. The ideal number to have for such a graph would be the percentage of subjects with any ARIA (either edema or hemorrhage), but this is not consistently reported. Typically, ARIA-E and ARIA-H percentages are reported separately and one cannot determine what percentage of subjects had either ARIA-E or ARIA-H. We settled on using the percentage of subjects with ARIA-E but note that this likely underestimates the amount of potential functional unblinding. Even with this caveat, there is a significant correlation across studies between percentage of ARIA-E and clinical effect (Figure 2). It seems likely that functional unblinding contributes importantly to the perceived benefit of anti-amyloid antibodies. It is also biologically plausible that ARIA-E is more frequent in trials with greater amyloid plaque removal and that the plaque removal, not the ARIA-E, accounts for the association with clinical effect. However, without cooperation from the pharmaceutical companies, or some coercion from the FDA, the field will not get a transparent view of how important a bias functional unblinding represents.

**Tau PET, a Better Biomarker, Is Unchanged by Donanemab**
Lastly, it is clear, as described above, that tau pathology is a far better correlate of cognitive status than amyloid. A drug that acts to correct a critical step in amyloid misprocessing and that has a true impact on longitudinal cognitive trajectories would be expected to impact NFT burden over time.   In fact, the first trial of immunotherapy targeting amyloid demonstrated that a treatment that removed amyloid plaque appeared to have no impact on tau pathology. The first active immunization study was halted in 2003 for several cases of encephalitis [25], but subjects were followed longitudinally and some came to autopsy. Several subjects who generated a robust antibody response were found to have a remarkable reduction of amyloid plaque, but they all died, nonetheless, with end-stage dementia and end-stage NFT pathology [26]. The interpretation was that successful vaccination with removal of amyloid plaques did not change NFT burden at death, but one might still hope that it had slowed the progression of tau pathology (something that could not be determined from postmortem assessments). With the advent of tau PET, we can now track the change in NFT burden over time. The donanemab study, reporting the largest effect size to date, included longitudinal tau PET scans on a large number of individuals (~400-600 per arm). In the analysis of change in tau PET (measured in two different regions-of-interest) over 76 weeks, there was no difference (or trend even) between active treatment and placebo [3].

**Conclusion**
In summary, (1) there is little to no evidence that the surrogate endpoint used for accelerated approval of anti-amyloid antibodies predicts clinical outcome within a trial, (2) the purported beneficial effects of anti-amyloid therapies may be driven by functional unblinding due to ARIA and (3) the lack of impact on tau pathology over 76 weeks argues strongly against clinically-relevant disease modification by donanemab.

These drugs are expensive and potentially dangerous. It is not clear that they provide any true clinical benefit. Their approval by the FDA and coverage by the Centers for Medicare and Medicaid Services may result in large uptake and considerable morbidity and even mortality due to brain edema and hemorrhage. Their widespread use would also make it challenging to run clean, well-powered clinical trials of more promising candidates. The field—broadly-defined to include physicians, regulatory personnel, patient advocacy groups, journal reviewers and editors—should demand more transparency and data-sharing from the pharmaceutical companies before the reported clinical outcomes are taken at face value.

**Conflicts of Interest**
Dr. Greicius is a co-founder and shareholder of SBGneuro.

**Funding Sources**
Dr. Winer is supported by the National Institute on Aging (F32AG074625).

**Acknowledgements**
The authors would like to thank Dr. Manisha Desai and Dr. Vivek Charu for their discussions regarding the statistical methods of the phase 3 donanemab clinical trial. The authors would also like to thank Dr. Kevin Lu for their insight into surrogate endpoints in cancer clinical trials.

**Figure 1**

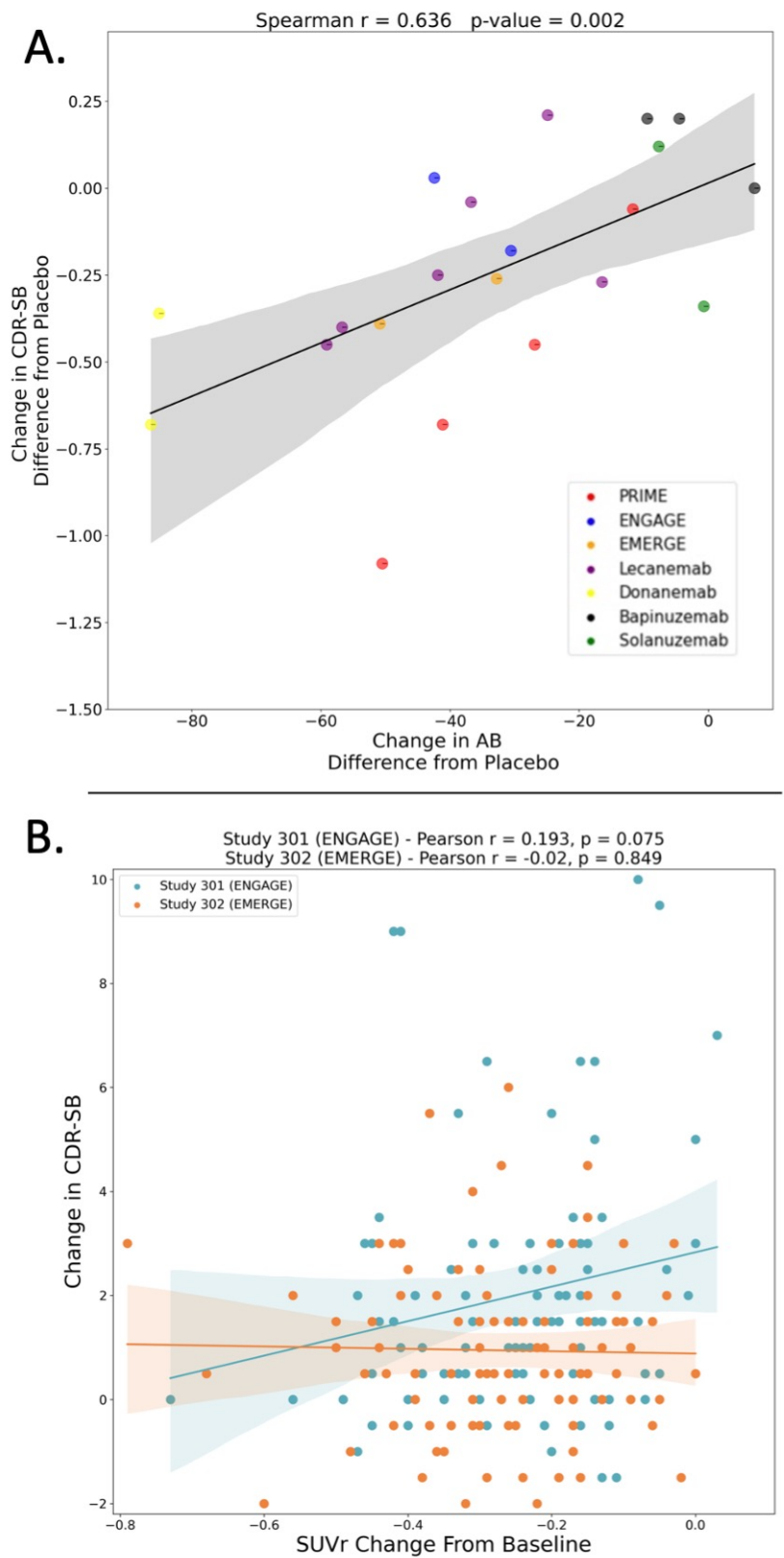

**Figure 1. (A) Relationship between amyloid reduction and cognitive outcomes** *across* **clinical trials (B) Relationship between amyloid reduction and cognitive outcomes** *within* **EMERGE and ENGAGE.** In Figure 1A, each point represents a clinical trial drug arm. For example, some trials have only one drug arm so the entire trial is represented by one point. However, for some trials, there were multiple drug arms (i.e., with different doses); in these cases, there will be more than one data point per clinical trial. Y-axis in Figure 1A represents change in CDR-SB (difference from placebo) for the drug group in each trial. X-axis in Figure 1A represents change in amyloid level (as estimated by change in centiloid, difference from placebo). For amyloid change, if data in the trials were reported as SUVr, we estimated their respective centiloid values using previously derived equations[27,28]. We also re-performed the analysis using percent amyloid change instead of centiloid change. This yielded similar results (Spearman r=0.64; p-value=0.003). In Figure 1B, each point represents a single study participant in the high-dose arm of EMERGE or ENGAGE. Figure 1B was constructed by extracting data from the aducanumab briefing document. The same graph is also included in the aducanumab FDA Clinical Pharmacology and Biopharmaceutics Review document [29]. The x-axis in Figure 1B represents change in amyloid SUVr. The y-axis in Figure 1B represents change in CDR-SB. The correlation in ENGAGE was weakly positive and nearly reaches significance, but notably of the two aducanumab trials this was the one with no statistically significant effect on CDR-SB. Abbreviations: AB: Amyloid, CDR-SB: Clinical Dementia Rating Scale - Sum of Boxes. SUVr: Standard uptake value ratio

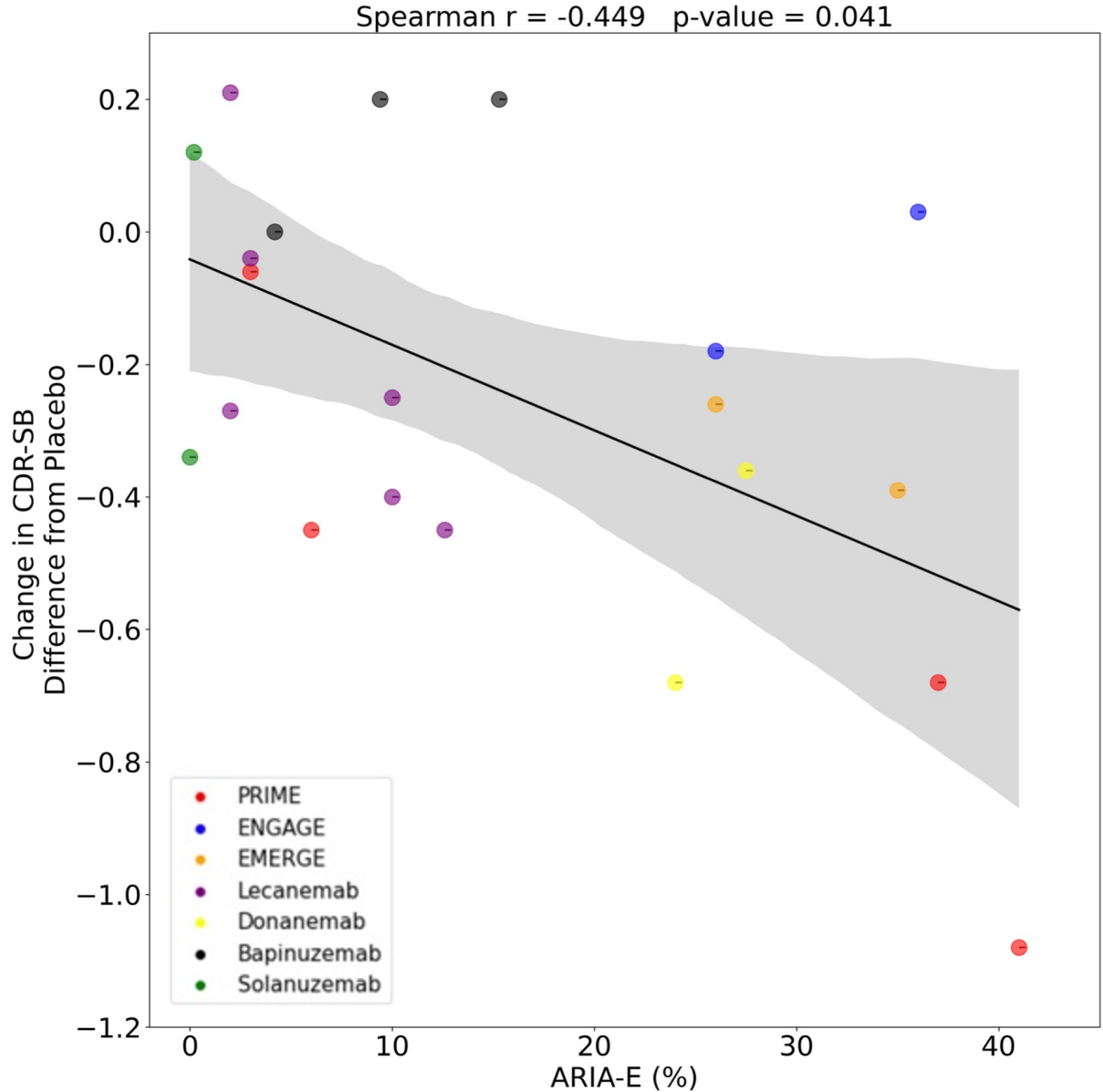

**Figure 2. Clinical trials with more ARIA-E are associated with better outcomes.** In the scatter plot, each point represents a clinical trial drug arm. As was the case for Figure 1A, some trials have only one drug arm so the entire trial is represented by one point. However, for other trials, there were multiple drug arms (i.e., with different doses); in these cases, there will be more than one data point per clinical trial. The x-axis represents percentage of participants in the respective drug arm that experienced ARIA-E. The y-axis represents change in CDR-SB (difference from placebo) for the drug group in each trial. Abbreviations: ARIA-E: Amyloid related imaging abnormality – edema; CDR-SB: Clinical Dementia Rating Scale - Sum of Boxes.